\documentclass[fleqn,usenatbib]{mnras}

\usepackage[T1]{fontenc}
\usepackage{ae,aecompl}

\usepackage{graphicx}	
\usepackage{amssymb}	

\title[Polarization and subpulse drifting of PSR B0031$-$07]{The drifting subpulses of PSR~B0031$-$07 and its synchronously modulated radio polarization}

\author[Ilie, Weltevrede, Johnston \& Chen]
{C.D. Ilie$^{1}$,
{P. Weltevrede$^{1}$}\thanks{email: patrick.weltevrede@manchester.ac.uk}, {S. Johnston$^{2}$} and {T. Chen$^{1}$}
\\
$^{1}$University of Manchester, Jodrell Bank Centre of Astrophysics, Alan Turing Building, Manchester, M13 9PL, United Kingdom\\
$^{2}$CSIRO Astronomy and Space Science, Australia Telescope National Facility, PO Box 76, Epping, NSW 1710, Australia\\
}

\date{Accepted XXX. Received YYY; in original form ZZZ}

\pubyear{2019}

\begin{document}
\label{firstpage}
\pagerange{\pageref{firstpage}--\pageref{lastpage}}
\maketitle

\begin{abstract}
We establish that for PSR~B0031$-$07 the orthogonal polarization modes switch at a single pulse level synchronously with the periodic drifting subpulses seen in total intensity. There are only four other pulsars known for which this phenomenon is observed. 
PSR~B0031$-$07 is unique as it is the only source in this group which has multiple stable drift modes. For both drift modes visible at our observing frequency centered at 1369~MHz, the modulation of polarization modes is synchronous with the drifting subpulses. 
In one of the drift modes, a discontinuity in the modulation pattern of polarization properties occurs halfway through the pulse, coinciding with a slight change in the slope of the intensity drift band. 
In contrast to what has been suggested for this pulsar in the past, this, plus other differences in the polarization of the modulated emission observed for the two drift modes, suggests that a drift mode change is more than a change in the underlying carousel radius and that magnetospheric propagation effects play an important role. 
The ellipticity evolves asymmetrically in time during the modulation cycle, which in the framework of a carousel model implies that the polarized sub-beams are asymmetric with respect to the sense of circulation, something which is not observed for
 other pulsars.
Birefringence in the magnetosphere, resulting in the orthogonal polarization modes to spatially separate, is not enough to explain these results. It is argued that more complex magnetospheric processes, which possibly allow conversion between orthogonal polarization modes, play a role. 
\end{abstract}




\section{Introduction}
\label{sect:introduction}

Significant progress has been made in understanding the mechanism responsible for the observed coherent radio emission of pulsars since their discovery more than 50 years ago \citep{hbp+68}. Nevertheless, due to the complexity of the  observed emission of pulsars, there is not yet a theoretical model which is able to account for all the diverse phenomenology. The work presented in this paper aims to examine the connection between the organised intensity variability seen in single pulses and the synchronous periodic switching of the polarization properties, in particular the position angle (PA), between two orthogonal polarization states. Implications for the radio emission mechanism are discussed.

\begin{figure}
\begin{center}
\includegraphics[width=0.9\hsize]{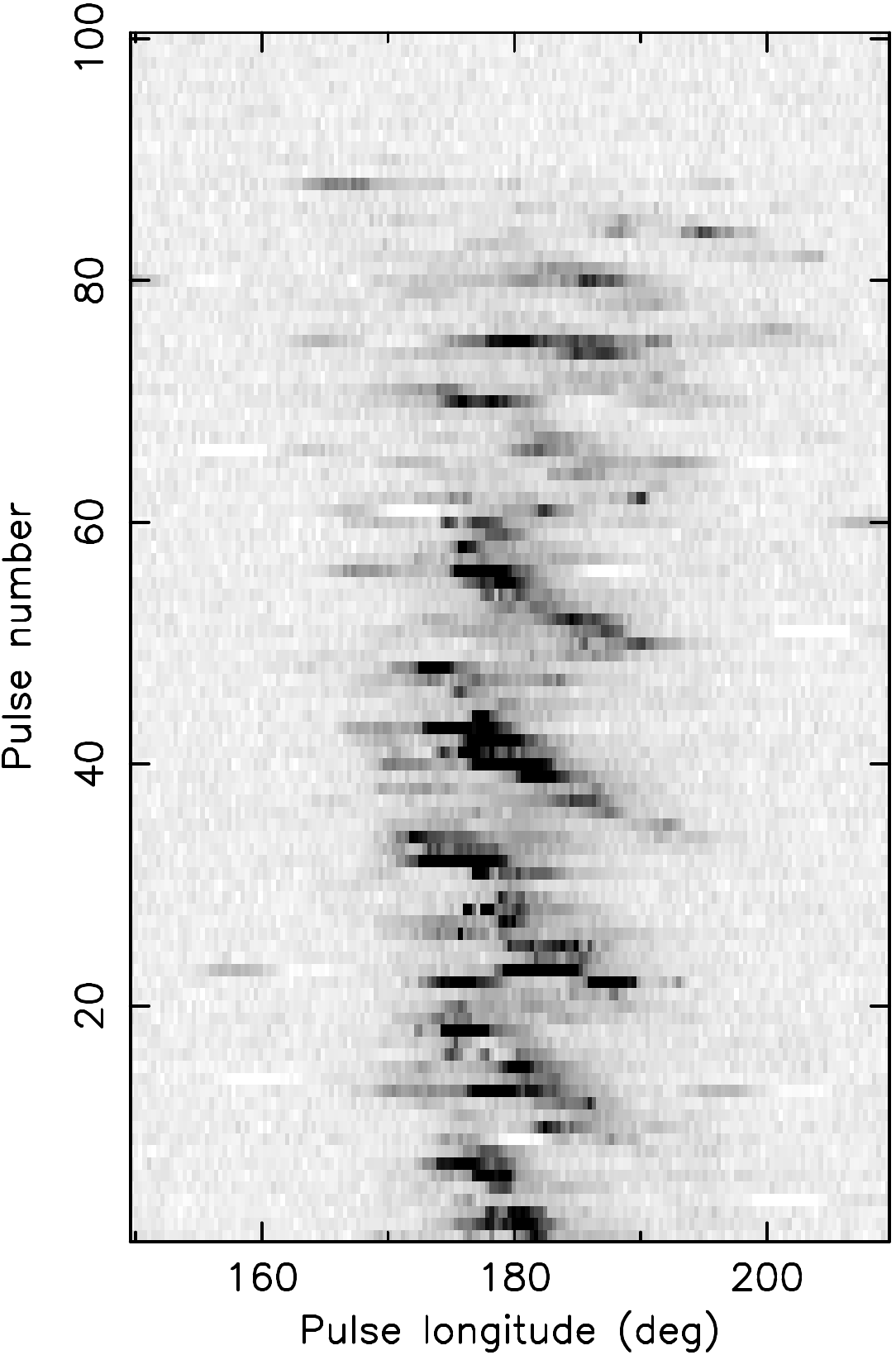}
\end{center}
\caption{A small interval of 100 pulses of PSR~B0031$-$07 halfway through the first observation  at a centre frequency of 1369~MHz. Drift mode A drift bands (up to pulse number 60) are followed by mode B drift bands and a null. In this figure the brightest features are saturated (the colour range covers $\sim$20\% of the dynamic range) to reveal weaker emission.}
\label{fig:stack}
\end{figure}

Although the individual pulses of pulsars are known to show chaotic changes in shape, many show systematic variations such that the emission is observed to shift in rotational phase in an organised manner, producing structures in the form of oblique intensity bands in the pulse-stack. This is illustrated in Fig.~\ref{fig:stack} for PSR~B0031$-$07 (J0034$-$0721). This phenomenon is referred to as \textit{drifting subpulses} and was first seen in pulsars by \citet{dc68}. The pattern can be defined with two periodicities: $P_2$, the spacing between two consecutive subpulses in rotational phase (or pulse longitude) and $P_3$, the separation between the diagonal drift bands in pulse number (or pulse periods $P$). 
Several authors \citep{wes06,wse07,bmm+16,bmm+19} have investigated the organised variability of single pulses and concluded that subpulse drifting is common in pulsars, being detectable in about half the pulsars investigated. Hence, the phenomenon is thought to be an essential component of the radio emission mechanism.

The most well established model to explain drifting subpulses is known as the \textit{carousel model} \citep{rs75}. In this model, the emission beam consists of smaller sub-beams which rotate around the magnetic axis due to an $\vec{E} \times \vec{B}$ drift force.
Individual subpulses correspond to these sub-beams and because of the motion of the carousel, each sub-beam progressively moves across the line of sight of an observer, resulting in the systematic shift as a function of pulse longitude. 
Since \citet{rs75}, many authors have either refined the carousel model (e.g. \citealt{dr99,gs03,wri03,vt12,hsw+13}) or suggested alternative models which could explain drifting subpulses (e.g. \citealt{cr04,gmm+05,fkk06}).

The PA of some pulsars trace an S-shaped swing during a stellar rotation. This is explained by the Rotating Vector Model (RVM) \citep{rc69,kom70} as the change in projection of the dipolar magnetic field lines on the plane of the sky. In a number of pulsars, the gradual S-shape is interrupted by discontinuities in the shape of $90\degr$ jumps in the PA. These are explained with the co-existence of two Orthogonally Polarized Modes (OPMs) of radiation \citep{mth75,brc76}.
The origin of these OPMs, and the reason why different OPMs can dominate in different parts of a pulse profile, are still uncertain. Some authors (e.g. \citealt{gan97}) suggested that the production of OPMs is intrinsic to the radio emission mechanism. Other authors (e.g. \citealt{mel79,am82,ab86}) suggested that the two OPMs are the two natural modes of propagation in the highly-relativistic magnetized plasma inside the pulsar magnetosphere. These two propagation modes are the Ordinary mode (O mode) and the eXtraordinary mode (X mode), which are predicted to be linearly polarized and orthogonal to each other.

For four pulsars it is observed that the dominant OPM periodically switches, and that this occurs synchronous with the drifting subpulses as observed in total intensity.
\citet{rrs+02} observed sharp $90\degr$ OPM transitions in the subpulses of PSR~B0809+74, which explained the periodic deviations from the RVM which were reported by \citet{thh+71} for that pulsar. \citet{rrs+02} showed that the flips between the dominant OPMs occur synchronous with the drifting subpulses. This finding was soon followed by the discovery of similar behaviour in PSR~B1237+25 \citep{rr03}, and in PSRs~B0320+39 and B0818$-$13 \citep{edw04}. For PSR~B1237+25 the strict synchrony of the OPM flips and the drifting subpulses is not established, since no technique such as $P_3$-folding (see Sect.~\ref{sect:p3fold}) was used,

In order to explain the fact that, for some pulsars, the switches between OPMs are modulated by the periodicity of drifting subpulses, the carousel model needed some revision.
\citet{rr03} suggested that a single circulating carousel system (i.e. the pattern of discharges in the polar gap) produces two ``images'' of emission, each correspond to one of the observed OPMs. An azimuthal offset between the two images (with respect to the magnetic axis)
would result in the observed periodic flips in the dominant PA.
The two images are believed to arise due to birefringence in the pulsar magnetosphere \citep{rrv+06}. This is expected since the X mode is not affected by refraction, thus it propagates in a straight path from the emitting region, while the O mode experiences refraction and therefore propagates on a curved path before escaping the pulsar magnetosphere (e.g. \citealt{pl00,pet01,lyu02,wsv+03}).
Hence, two images of the carousel will be produced; each image has mutually
orthogonal polarization and could be magnified differently.

In this work, we show that modulation of OPMs synchronous with the drifting subpulses exists for PSR~B0031$-$07. This is a particularly interesting and widely studied pulsar, as it displays drifting subpulses with three different stable drift modes. These are modes A ($P_3 \simeq13P$), B ($P_3 \simeq7P$) and C ($P_3 \simeq4P$) \citep{htt70}, where $P_3$ is found to be variable within a given mode as well \citep{mbt+17}.
The detectability of these modes depends on the observing frequency \citep[e.g.][]{smk05,sms+07}. For example, at 1.4~GHz only modes A and B have been observed \citep{sms+07}. \cite{man75} investigated the polarization properties of this pulsar and reported the presence of two OPMs.
Our investigation builds on an initial analysis presented in a Masters thesis by \citet{chen15} and preliminary results were published in \cite{iw18}. However,  detailed analysis was hindered by instrumental variations on timescales comparable with the pulsar period $P$. In this analysis, we describe the observed behaviour in more detail using new data and discuss potential interpretations.

The structure of this paper is as follows. Sect.~\ref{sect:observations3} describes the observations of PSR~B0031$-$07. In Sect.~\ref{sect:driftmodes}, it is explained how the data is separated in multiple drift modes and fluctuation analysis is used (Sect.~\ref{sect:fluctuation}) to confirm that this is done successfully. The polarization properties of this pulsar are discussed in depth in Sect.~\ref{sect:polarization} before in Sect.~\ref{sect:p3fold} it is established that the observed OPMs switch synchronously with the drifting subpulses.
The results are discussed in Sect.~\ref{sect:discussion3} and the conclusions are presented in Sect.~\ref{sect:conclusions3}.

\section{Observations}
\label{sect:observations3}

Two observations of PSR~B0031$-$07 are described in this paper, which were performed using the H-OH receiver of the Parkes  telescope with a bandwidth of 256~MHz, split into 512 frequency channels, centred at 1369~MHz. The data was recorded with the backend system known as the Parkes Digital FilterBank PDFB4. 
The first observation was made on 2016 April 9 and the second on 2016 August 29.
In the two observations 13,622 and 7,951 individual pulses were recorded respectively.

Data was de-dispersed, de-Faraday rotated and folded off-line using \textsc{dspsr} \citep{vb11} which allows the data to be re-organised in the form of a pulse-stack. After folding, the time resolution was 1024 pulse longitude bins per stellar rotation. The individual frequency channels were averaged, excluding those affected strongest by radio frequency interference (RFI). Some broad-band RFI was present in the data, including periodic bursts of RFI believed to be associated with a pump. Given the large number of pulses recorded, the effect on the analysis was found to be minimal.
At the start of each observing session, a pulsed calibration signal was recorded for two minutes, while the telescope was pointed offset from the pulsar. This allows correction for differential gain and phase, which was applied to the data using the \textsc{psrchive}\footnote{http://psrchive.sourceforge.net/} software package.

The two observations were combined to improve the sensitivity. The rotational-phase alignment was achieved using a rotational ephemeris which was optimised for the rotation period $P$ and its time derivative $\dot{P}$ based on observations from the 76 m Lovell radio telescope in Jodrell Bank. 
Although the analysis presented in this paper is based on the combined observation, it was checked that the results are consistent with those obtained from the two individual observations.

\begin{figure}
\begin{center}
\includegraphics[width=0.99\hsize]{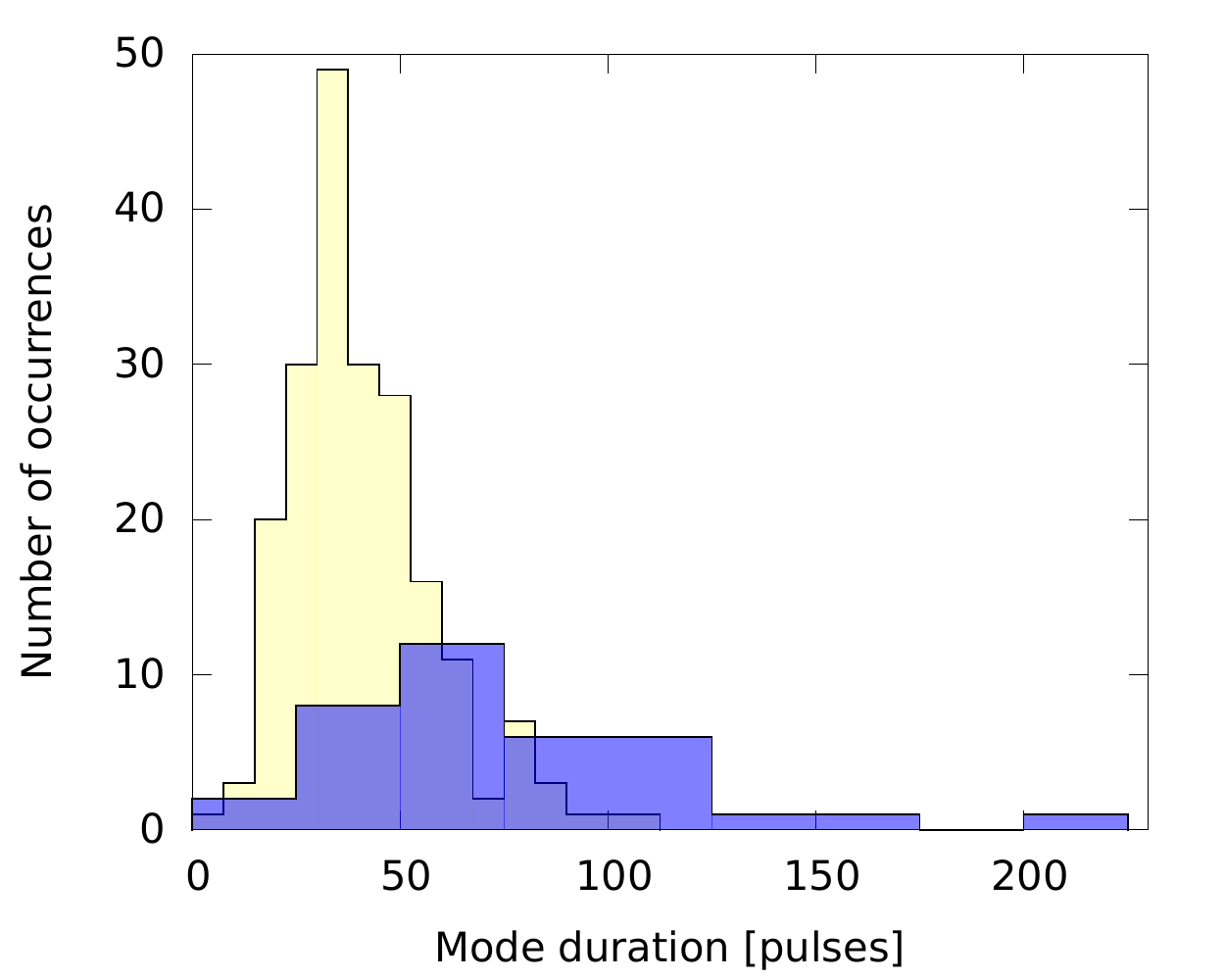}
\end{center}
\caption{Distributions of the duration of stretches of data in a certain drift mode. Drift modes A and B correspond to the light and dark colored distribution respectively (yellow and blue in the online version).}
\label{fig:mode_length}
\end{figure}

\begin{figure*}
\begin{center}
\includegraphics[width=0.3\hsize]{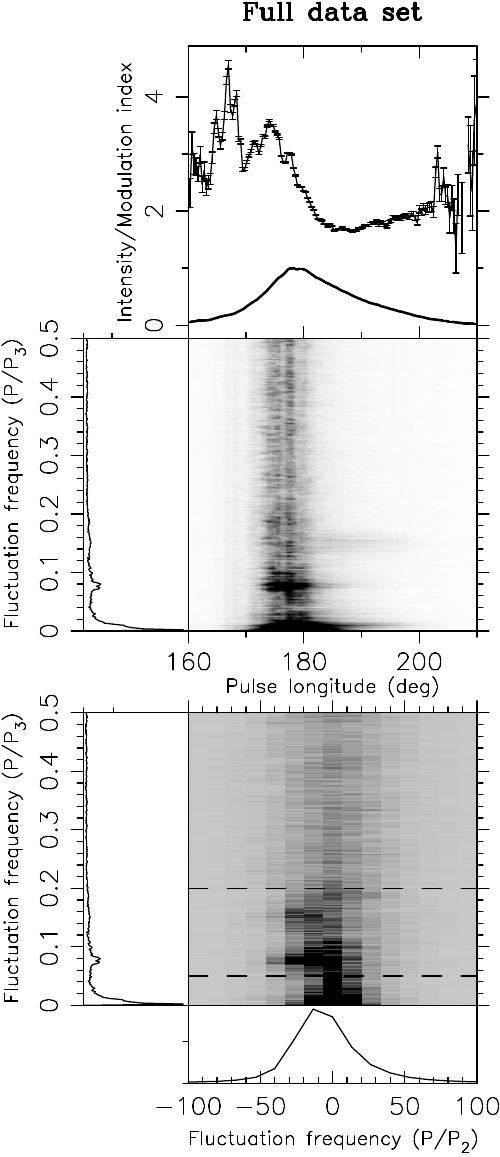}\hspace{0.04\hsize}
\includegraphics[width=0.3\hsize]{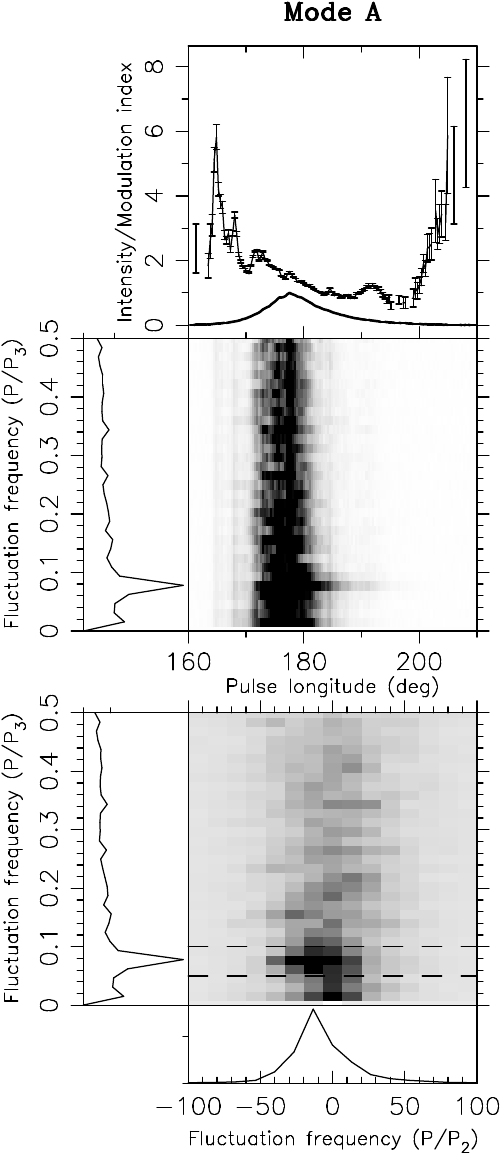}\hspace{0.04\hsize}
\includegraphics[width=0.3\hsize]{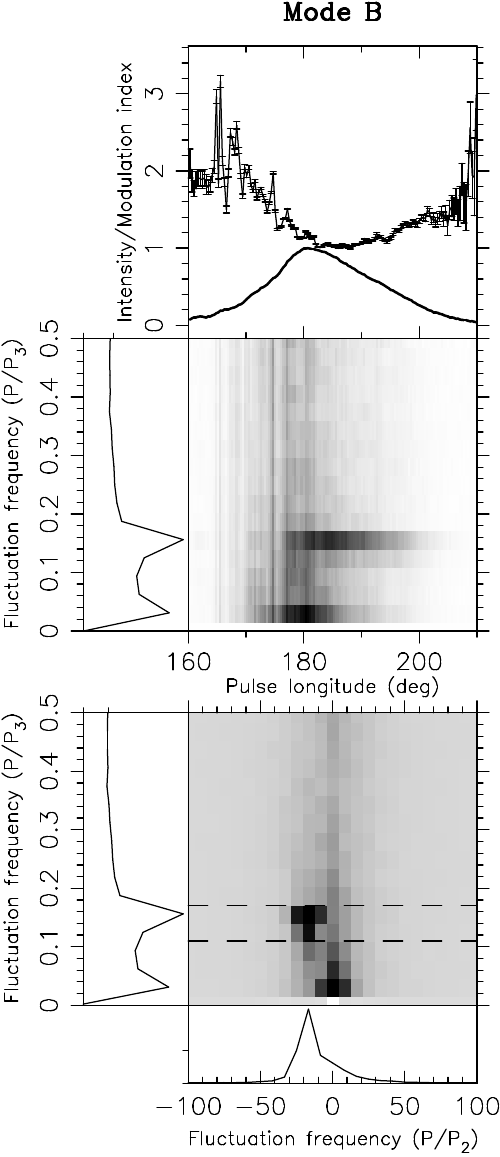}
\end{center}
\caption{The results of a fluctuation analysis of PSR B0031$-$07. The top panels show the integrated pulse profiles, as well as the modulation index (data points with their associated uncertainties if its significance exceeds $3\sigma$). The middle panel displays the LRFS with the horizontally integrated power shown in the side panel. The bottom panels show the 2DFS, with two side panels, which display the horizontally and (between the two dashed lines) vertically integrated power respectively. The brightest features in the grey-scale plots are saturated to highlight the weaker features.
\textit{Left:} The not-mode-separated data. 
\textit{Middle:} Drift mode A. 
\textit{Right:} Drift mode B.}
\label{fig:fs}
\end{figure*}

\section{Analysis}
\label{sect:analysis3}

All analysis in this section was performed with
\textsc{psrsalsa}\footnote{https://github.com/weltevrede/psrsalsa}
\citep{wel16}.

\subsection{Drift mode identification}
\label{sect:driftmodes}

A small interval of the pulse-stack of PSR~B0031$-$07 containing 100 consecutive pulses is displayed in Fig.~\ref{fig:stack}, showing several mode A drift bands up to pulse number 60. After this, shallower mode B drift bands can be seen followed by a null state, during which the radio emission mechanism appears to be inactive.
As noted by for example \citet{smk05}, the pattern of drifting subpulses is clearer at lower observing frequencies. Nevertheless, the choice was made to observe at this frequency because of the excellent performance of the Parkes telescope in producing polarization calibrated pulsar data (see e.g. \citealt{wj08,jk18}).

As we are interested in the modulation of the OPMs, the analysis which follows requires  the data to be separated in the different drift modes, since each drift mode has different polarization and modulation properties.
The adopted methodology of mode separation is as follows. The pulse-stack was visually inspected.
One occurrence of mode C was seen in the pulse-stack (the length of the whole observation is $\sim5$ hours), as several drift bands following an occurrence of mode B. However, due to the rarity of mode C, we did not attempt to include this mode in any further analysis. 
Each individual pulse was assigned one of the following categories: mode A, mode B, mode C, null state, or the emission was too weak to be classified. The last category was included in order to avoid possible mis-labeling weak emission. There were no individual pulses where RFI affected the mode identification.

Out of a total of 21,573 individual pulses, the pulsar was in a null state for 8,606 pulses (39.9$\%$) and the pulsar was on (in one of the three drift states) for 43.1$\%$ of the time (9,294 pulses). For the rest of the pulses, emission was observed, but too weak to be classified reliably (17.0$\%$).
During the time the pulsar was on and classifiable, 30.5$\%$ of the time it was in mode A (2,838 pulses), 69.2$\%$ of the time it was in mode B (6,430 pulses) and 0.3$\%$ of the time it was in mode C (26 pulses).
At 328~MHz, \citet{smk05} reported that PSR B0031$-$07 was on (in one of the three drift modes) 61.8$\%$ of the time (of which 17.8$\%$ mode A, 80.1$\%$ mode B and 2.1$\%$ mode C). 
This confirms that mode A is more common at higher radio frequencies \citep{smk05,sms+07}.
The null fraction as reported by \citet{smk05} is very similar. 

Distributions of the duration of individual stretches of mode A and B data are shown in Fig.~\ref{fig:mode_length}. From these plots, it can be seen that bursts of mode A emission tends to last longer compared to mode B emission. The mode A bursts last between $\sim$25 and $\sim$200 pulses (between 2 and 15 consecutive drift bands). Mode B bursts range from a few pulses up to 100 pulses (between 1 and 15 consecutive drift bands).

Mode separation is somewhat subjective because of variability in the emission unrelated to drifting subpulses. Nevertheless, a fluctuation analysis on the mode-separated data confirms that the mode separation was successful.
The results of the fluctuation analysis on the total intensity both before and after mode separation are shown in Fig.~\ref{fig:fs}. A detailed description of this figure is provided in the following subsections.

\begin{figure*}
\begin{center}
\includegraphics[width=0.325\hsize]{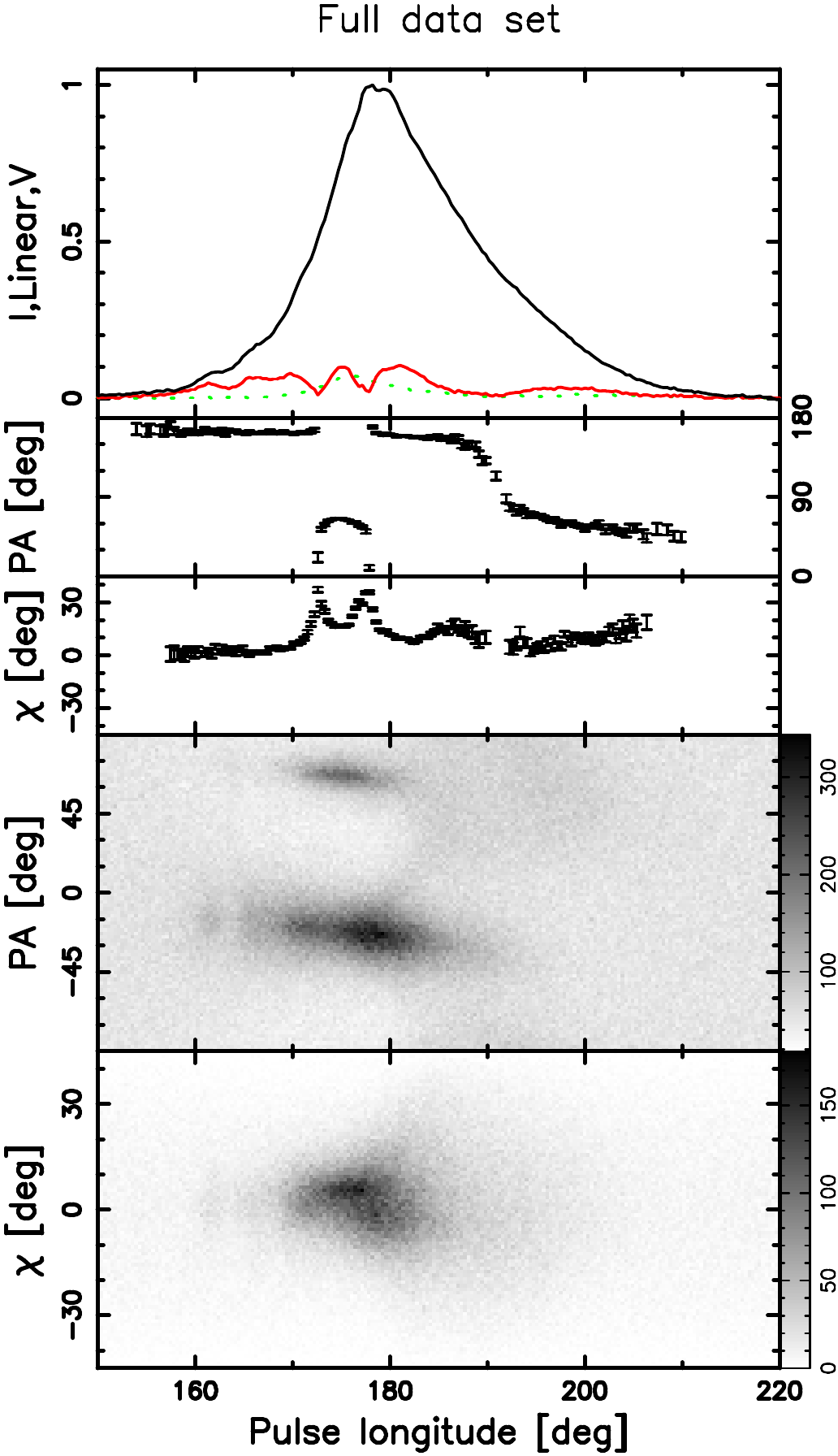}\hspace{0.005\hsize}
\includegraphics[width=0.325\hsize]{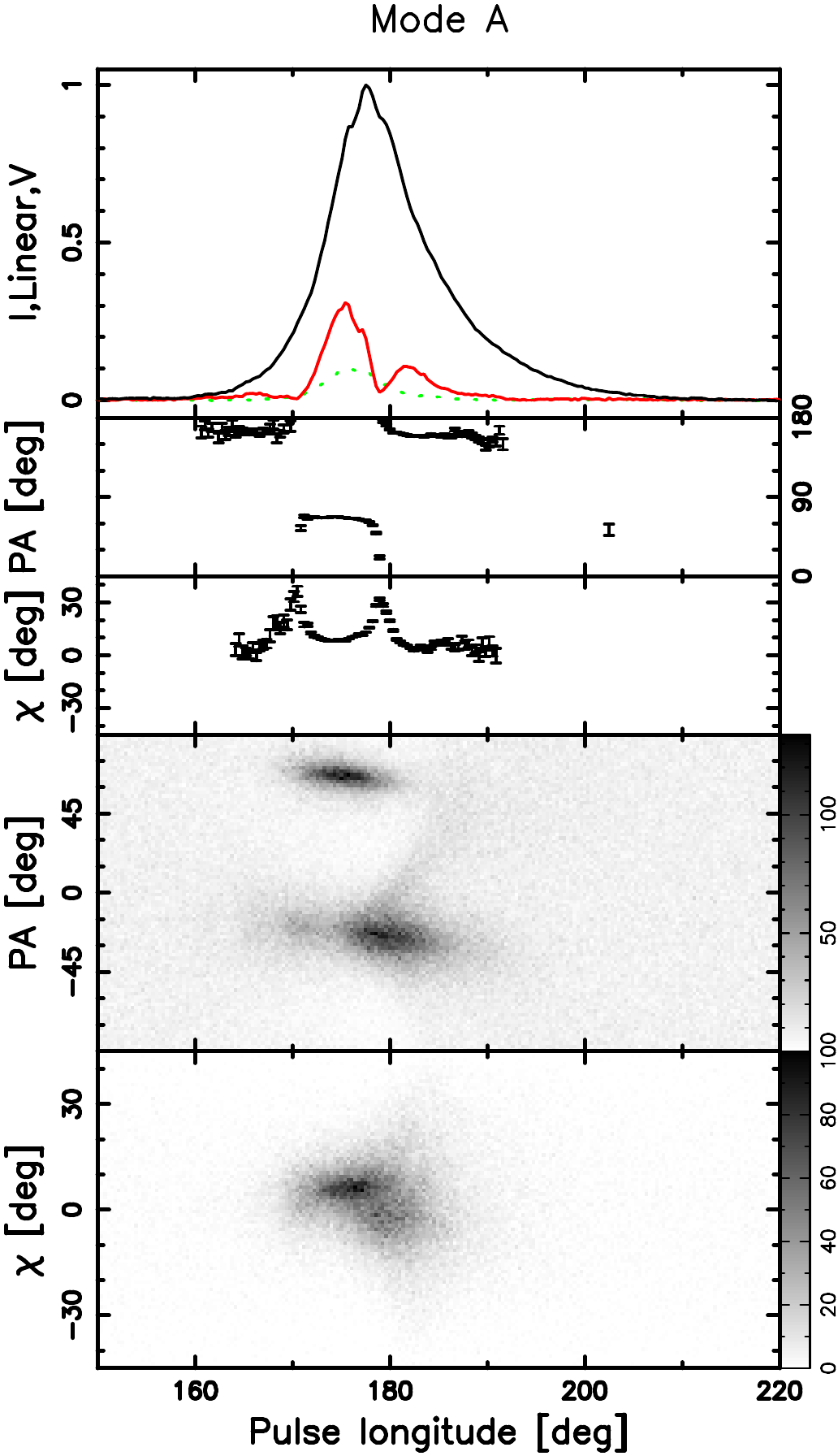}\hspace{0.005\hsize}
\includegraphics[width=0.325\hsize]{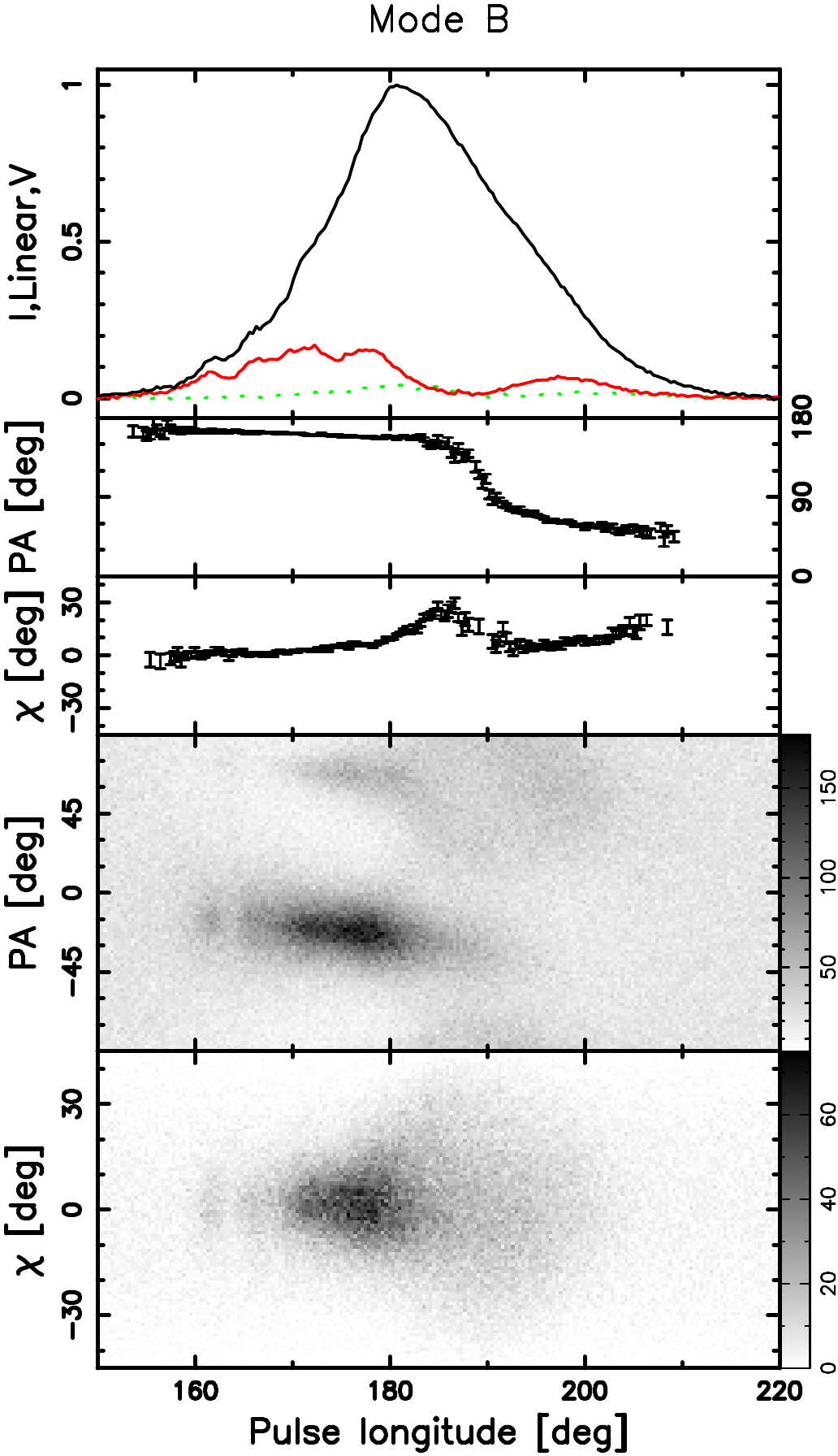}
\end{center}
\caption{Polarization properties of PSR~B0031$-$07. In the top panels, the total intensity pulse profile (black curve), linear polarization (red curve in the online version), and circular polarization (dotted curve which is green in the online version) are shown. The time averaged PA and ellipticity angle $\chi$ are shown in the second and third row of panels respectively.
The bottom two rows of panels display the PA and $\chi$ distribution obtained for individual pulses when the significance of respectively $L$ and $\sqrt{Q^2+U^2+V^2}$ exceeded 1$\sigma$ (the required significance for the average PA and $\chi$ was 2$\sigma$). Note that the average PA swing is mapped between $0\degr$ and $180\degr$, while the PA distribution covers $-90\degr$ to $90\degr$.
\textit{Top:} The not-mode-separated data. 
\textit{Bottom left:} Mode A data. 
\textit{Bottom right:} Mode B data.}
\label{fig:PA_all}
\end{figure*}

\subsection{Fluctuation analysis}
\label{sect:fluctuation}

All fluctuation analysis described in this subsection was done in the spectral domain. See \citealt{wes06,wse07} for details about the methodology.

The modulation index quantifies how much the intensity varies from pulse to pulse for a certain  pulse longitude and is equal to the standard deviation of the intensity divided by its average. The modulation index, along with the pulse profile, is shown in the top panels of Fig.~\ref{fig:fs}. This reveals an asymmetry such that the leading part of the profile is more strongly modulated, and that the two drift modes have different modulation properties. To quantify the periodicities in the data, and their difference for the two modes, fluctuation spectra are calculated.

The computation of the Longitude Resolved Fluctuation Spectrum (LRFS; \citealt{bac70}) allows the investigation of the repetition frequency of the drifting subpulse pattern $P_3$. The LRFS is obtained by averaging the power spectra obtained from blocks of pulses. For the not-mode-separated data large blocks of 512 pulses were used, which gives a high spectral resolution. For the mode-separated data, the choice of block size was made based on the typical mode length as inferred from Fig.~\ref{fig:mode_length}. Since the pulses in each block must be continuous, a large block size implies limited data can be analysed. The mode A and B separated data was organised into sequences of 64 and 32 consecutive pulses respectively. Care was taken to only consider blocks in which all pulses were classified to be in the same mode and incomplete blocks were discarded from the analysis.

The LRFS for modes A and B and the non-mode-separated data are shown below the pulse profiles in Fig.~\ref{fig:fs}. The left-most LRFS in Fig.~\ref{fig:fs}, which is based on the not-mode-separated data, reveals three main features. 
The first is centred at zero fluctuation frequency, pointing towards fluctuations with long timescales, and is likely to be associated to nulling and drift mode changes.
There is a strong feature at a fluctuation frequency $P/P_3\simeq0.07$ cycles per period (cpp), indicating $P_3 \simeq P/0.07 \simeq 14P$ which corresponds to drift mode A.
The fact that the feature is resolved in fluctuation frequency indicates that $P_3$  is variable. A weaker feature corresponding to mode B, which is more pronounced in the trailing half of the profile, can be seen at $\sim0.16$ cpp or at $P_3 \simeq 6P$.

As a confirmation for the successful mode separation, in  analysis of the  mode-separated data (middle and right column in Fig.~\ref{fig:fs}), only the periodicity of the drift mode of interest remains. This is most pronounced in the side panels, which show the horizontally integrated power.

To measure $P_2$, or demonstrate the existence of drifting subpulses, the 2-Dimensional Fluctuation Spectrum (2DFS, \citealt{es02}) can be used. In the 2DFS (bottom row of panels in Fig.~\ref{fig:fs}) there is power at the same $P/P_3$ frequencies as observed in the LRFS.
 The fact that the power is offset horizontally from zero, resulting in a negative $P_2$, means that, as expected, the subpulses drift to the leading side of the profile (see Fig.~\ref{fig:stack} as well). $P_3$ and $P_2$ were measured for the mode-separated data by calculating the centroid of power the in the 2DFS (taking care to account for systematic uncertainties, see \citealt{wes06}).
For mode A, this results in $P_3 = (13.0\pm0.2)P$ and $P_2 = {\left(30^{+15}_{-10}\right)}\degr$.
For mode B, $P_3 = (6.9\pm0.2)P$ and $P_2 = {\left(20^{+7}_{-5}\right)}\degr$.
These values are consistent with Fig.~\ref{fig:stack} and measurements by \citet{sms+07} at 1167~MHz, as well as $P_3$ measurements by \citet{htt70} and \citet{smk05}. The measurements are consistent with $P_2$ being identical for the different drift modes, as was noted by \citet{htt70}.

\begin{figure*}
\begin{center}
\includegraphics[width=0.49\hsize]{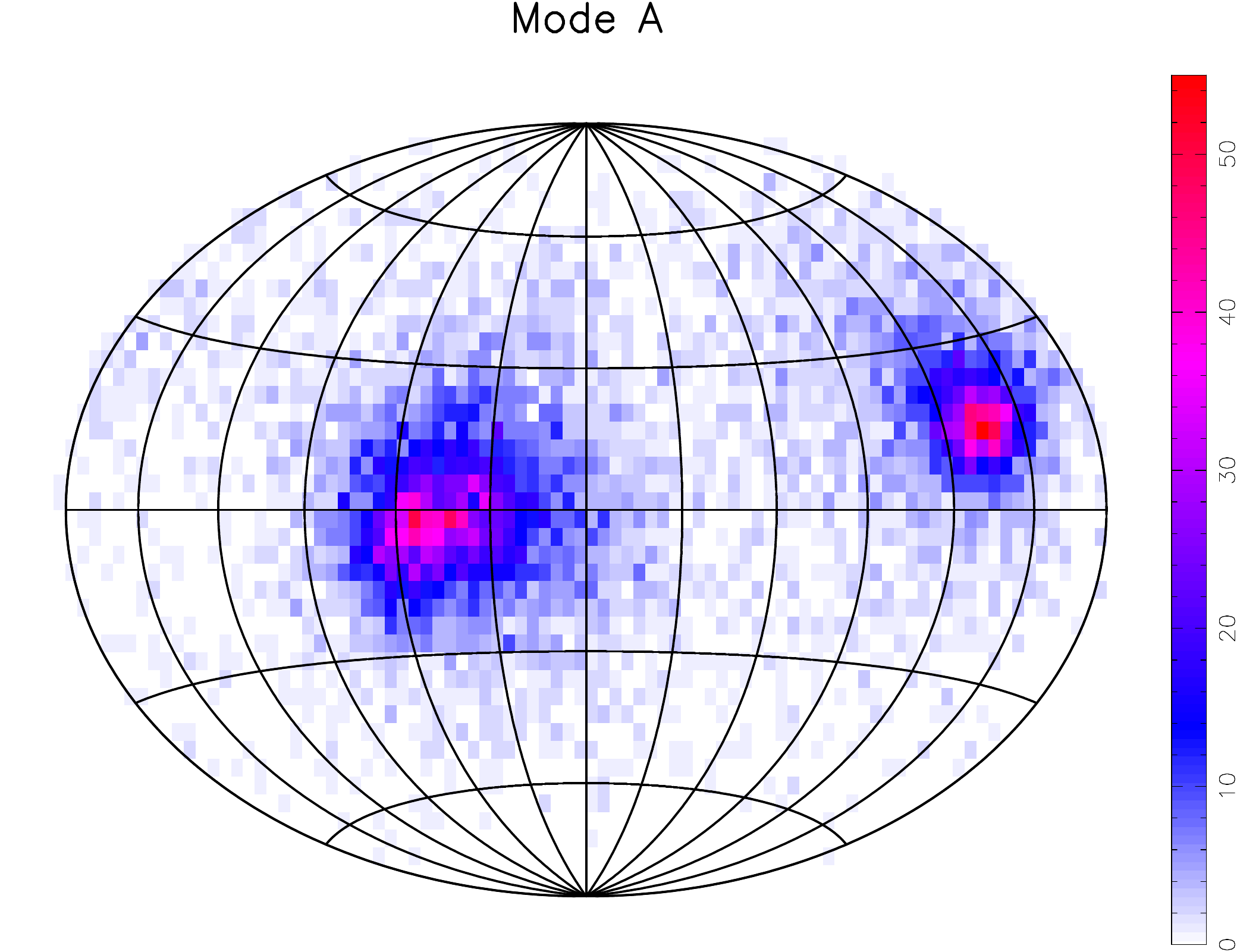}\hspace{0.01\hsize}
\includegraphics[width=0.49\hsize]{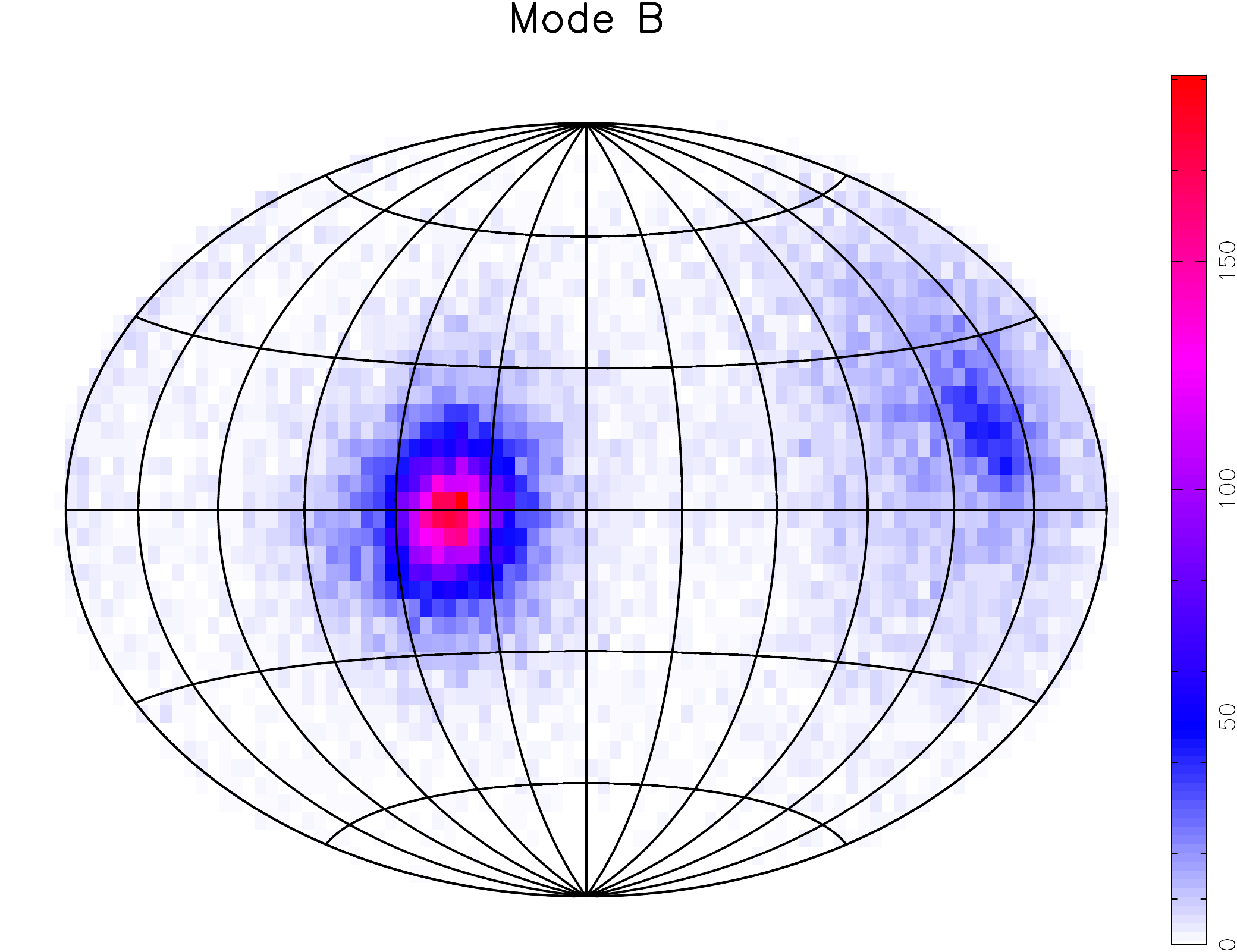}
\end{center}
\caption{Distribution of polarization orientations on a Poincar\'e sphere projected using the Hammer equal area projection. Only emission with both $L$ and total polarization exceeded a significance 1$\sigma$ were included, and samples from all pulse longitudes in the pulse window were considered, after re-binning the data to 64 longitude bins per rotation. The effect of the PA-swing was compensated for to reduce smearing (see text). The latitudes represent great circles of constant $\chi$ separated by $30\degr$ with pure linear polarization being the equator. The north and south poles correspond pure circular polarization with a positive and negative Stokes $V$ respectively. The meridians represent lines of constant PA, ranging from $-90\degr$ (left) to $90\degr$ (right), in steps of $15\degr$. 
\textit{Left:} Mode A data. 
\textit{Right:} Mode B data.}
\label{fig:poincare}
\end{figure*}

\begin{figure*}
\begin{center}
\includegraphics[width=0.4\hsize]{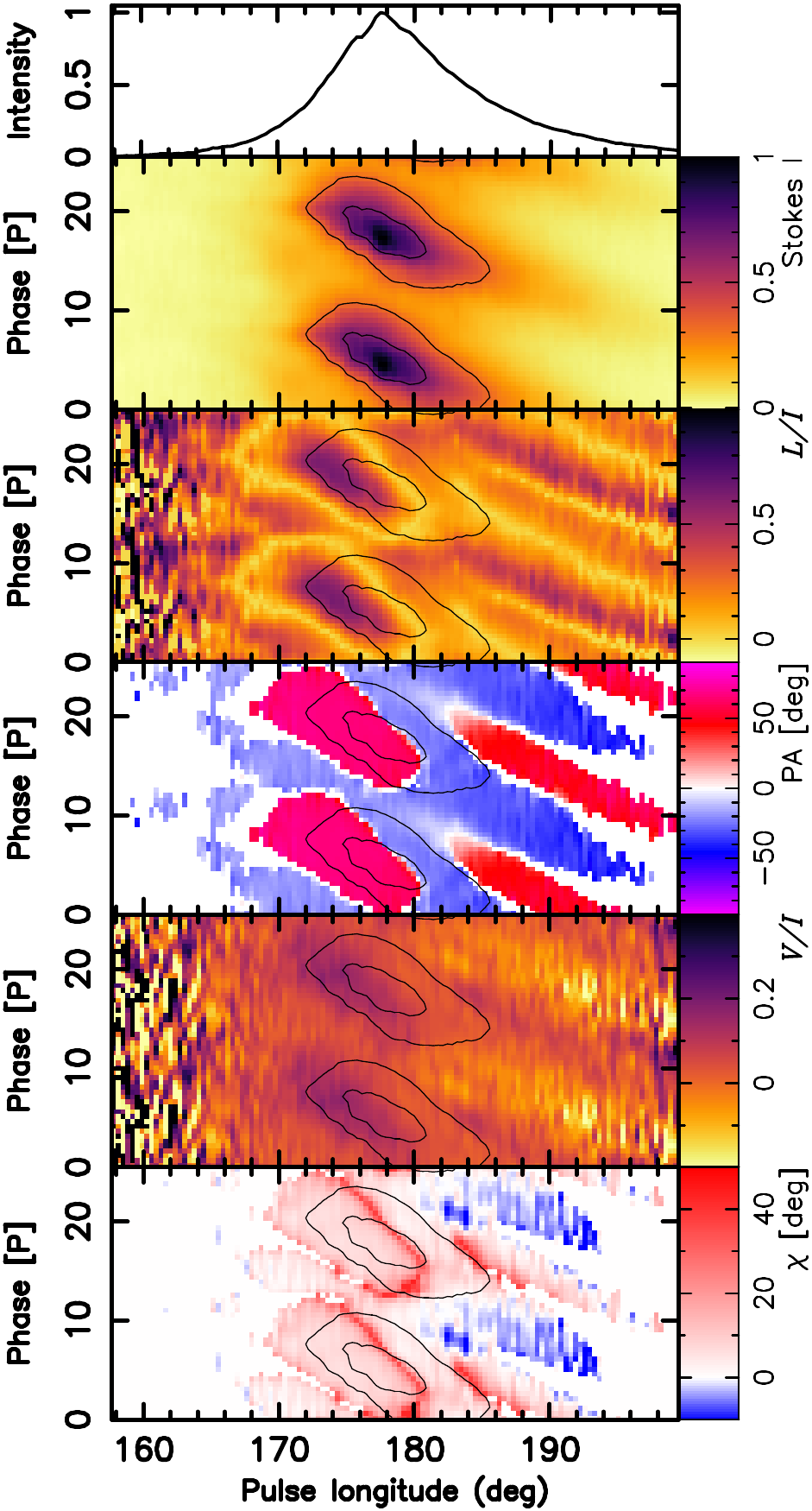}\hspace{0.1\hsize}
\includegraphics[width=0.4\hsize]{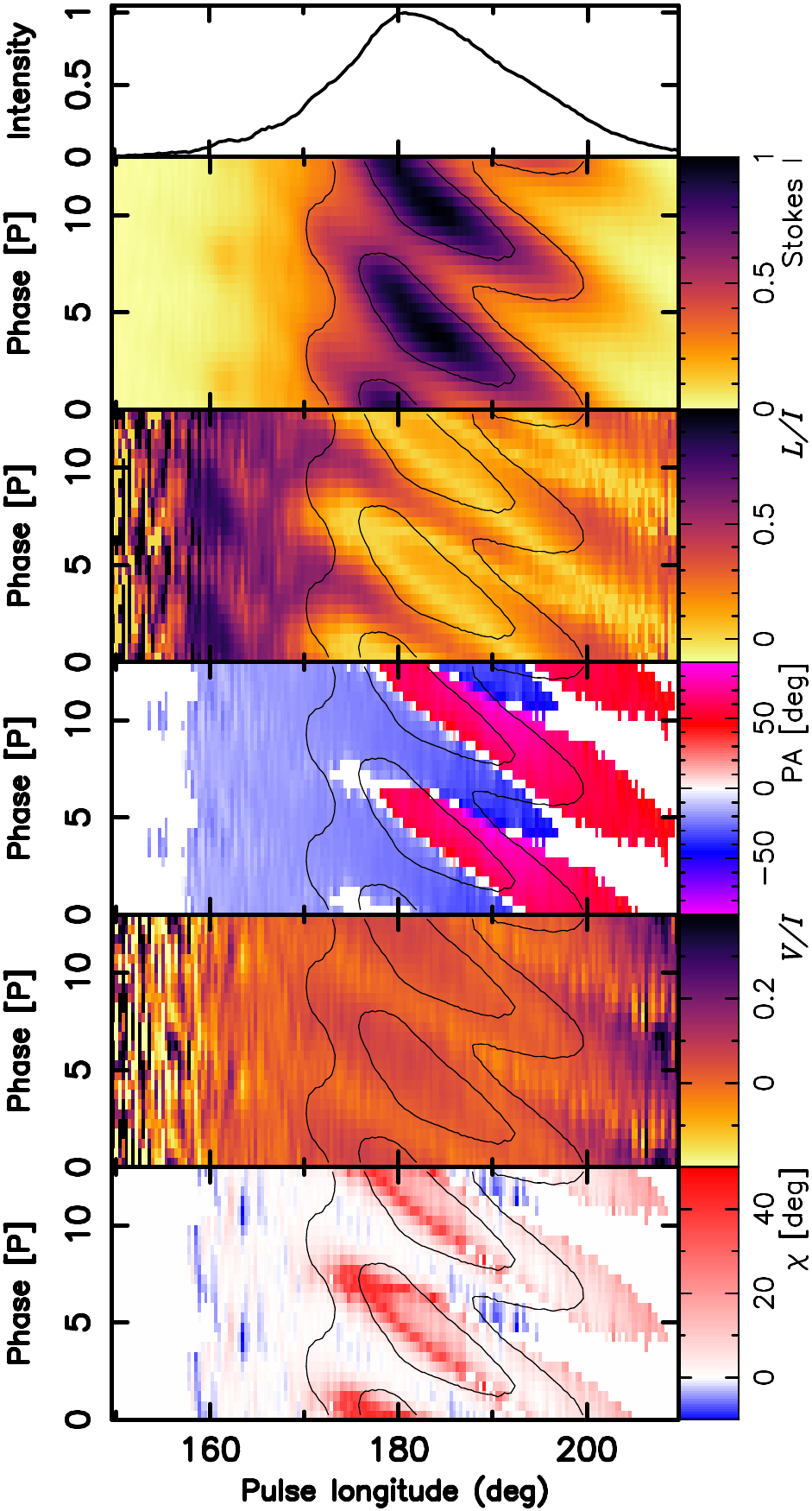} 
\end{center}
\caption{$P_3$-folds of drift mode A ({\em left}) and B ({\em right}) for PSR~B0031$-$07 are shown below the pulse profile. From top to bottom the $P_3$-folds correspond to Stokes $I$, $L/I$, PA, $V/I$ and $\chi$. The modulation cycle is shown twice for continuity. The contours, which are identical in each panel, correspond to the drift band shape observed in Stokes $I$. The PA and $\chi$ are shown if the significance of respectively $L$ and $\sqrt{Q^2+U^2+V^2}$ exceeds 3$\sigma$.}
\label{fig:p3fold_modeA}
\end{figure*}

\subsection{Polarization properties}
\label{sect:polarization}

Looking at the top panels of Fig.~\ref{fig:PA_all}, it is clear that modes A and B have significantly different intensity profiles and average polarization properties. The intensity profile of mode B peaks $\sim5\degr$ later in pulse longitude and the widths are also different. The measured 10$\%$-widths of mode A and B are $28\degr$ and $44\degr$ respectively with a nominal error of $1\degr$.
This shift in peak location of the pulse profile, and the variation in width, have been observed at other frequencies as well (e.g. \citealt{vj97,wf81,vj97,sms+07}).

The average linear polarization $L=\sqrt{Q^2+U^2}$ and circular polarization $V$ of the not-mode-separated data are low (where $Q$, $U$ and $V$ are Stokes parameters), as seen in the top panels of Fig.~\ref{fig:PA_all}. In comparison, $L$ in the mode-separated data is stronger, which suggests that different OPMs dominate for different drift modes, thereby causing depolarization when summed. 
The PA swings of the two drift modes are indeed different (see second row panels of Fig.~\ref{fig:PA_all}). The two OPM jumps observed in the PA swing of the more strongly polarized mode A do not occur for mode B. The occurrence of OPM jumps in PSR B0031--07 is known to be frequency dependent \citep{sms+07}.
The ellipticity angle $\chi$, defined via $\tan(2\chi)=V/L$, appears to peak where the two OPM transitions occur for mode A (see third row of panels of Fig.~\ref{fig:PA_all}). For mode B, the ellipticity angle is small, except around longitude $\sim189\degr$ and towards the trailing edge of the profile.

To further investigate the existence of OPMs, the PA distribution of individual pulses was determined. The obtained distributions (see fourth row of panels in Fig.~\ref{fig:PA_all}) reveal that at most pulse longitudes there is a mixture of OPMs observed for both modes A and B. We define the OPM observed at  PA~$\simeq 65\degr$ as OPM$_\mathrm{A}$, and the OPM at PA~$\simeq -25\degr$ as OPM$_\mathrm{B}$. 
For drift mode B, OPM$_\mathrm{B}$ dominates up to pulse longitude $\sim190\degr$. Compared to drift mode B, OPM$_\mathrm{A}$ dominates much more frequently in the individual pulses of drift mode A. This is especially true between pulse longitudes $\sim170\degr$ and $\sim180\degr$, thereby explaining the jumps in the integrated PA swing. 
Since there is a mixture of OPMs, depolarization of the integrated profile can be expected, as is observed.

In the PA distribution of drift mode A (fourth panel in the middle column of Fig.~\ref{fig:PA_all}), a  structure can be seen flaring off upwards from OPM$_\mathrm{B}$, starting at pulse longitude $178\degr$ and connecting with the other OPM at longitude $\sim190\degr$. Clearly the polarization state of the radiation is more complex than an incoherent summation of two OPMs. This structure is only visible in the drift mode A data. 

Unlike the drift mode B data, the mode A data has an ellipticity angle distribution (bottom row of panels in Fig.~\ref{fig:PA_all}) which does appear to be bi-modal, with one component being significantly more linear than the other. The origin of the bi-modality is revealed by exploring the distribution of polarization orientations which can be visualised in the form of a Poincar\'e sphere (Fig.~\ref{fig:poincare}). The data was first re-binned into 64 pulse longitude bins, in order to increase the significance of each sample. Since the peaks in the PA distributions in Fig.~\ref{fig:PA_all} shift only slightly in PA as function of pulse longitude, especially where the polarized emission is strongest, the introduced depolarization because of rebinning is minimal. Nevertheless, to compensate for this effect, the Stokes $Q$ and $U$ parameters for each sample were rotated before rebinning such that the time-averaged PA-swing was removed from the data. This was done for each drift-mode separated dataset separately. After rebinning, the time-averaged PA-swing as measured from the re-binned data was re-introduced before producing the Poincar\'e spheres as shown in Fig.~\ref{fig:poincare}.
The two OPMs manifest themselves as two islands offset by $\sim180\degr$ in longitude in the Poincar\'e sphere. OPM$_\mathrm{A}$ is clearly elliptical (appearing at the right hand side), while OPM$_\mathrm{B}$ is almost entirely linearly polarized.

The shown Poincar\'e spheres demonstrate that the two OPMs have polarization properties which are essentially independent on drift mode. OPM$_\mathrm{A}$ is clearly elliptical with $\chi\simeq8\degr$, while OPM$_\mathrm{B}$ is practically linear. This means that the two polarization modes are not situated on antipodal points on the Poincar\'e sphere. According to most theories, OPMs are expected to be orthogonal in both PA and ellipticity (e.g. \citealt{mmk+06}), however deviations from orthogonality have been previously observed for other pulsars too (e.g. \citealt{esv03,edw04}).

\subsection{Periodic polarization variability}
\label{sect:p3fold}

 As is also explained in \cite{iw18}, $P_3$-folding can reveal  the relation between the two OPMs and the periodicity of drifting subpulses.
$P_3$-folding is the process where data is averaged over the modulation cycle which results in the average shape and properties of a drift band. Since $P_3$ is not fixed, but varies, as demonstrated in Sect.~\ref{sect:fluctuation}, these variations need to be take into account when averaging successive drift bands over large stretches of data (e.g. \citealt{dr01,vkr+02}). Here the variability was compensated for using a tool from the \textsc{psrsalsa} software package, a tool which was first used by \citet{hsw+13}.

The folding algorithm works as follows. For both drift modes, the $P_3$ cycle is resolved in $2P_3/P$ bins.
 This oversampling by a factor of two implies that subsequent bins are dependent, but this gives a smoother visual result. The effective resolution was made to match the intrinsic resolution by smoothing the data using a Gaussian
smoothing kernel during the folding process.
Before being folded, the  mode-separated data was split in blocks of 3$P_3$ pulses.
Care was taken to only use blocks containing continuous stretches of pulses in a given mode, similar to what was done for the fluctuation analysis described in Sect.~\ref{sect:fluctuation}. 
Each block was first folded at a fixed value of $P_3$, taken to be the average value for the given drift mode as determined from the fluctuation analysis described in Sect.~\ref{sect:driftmodes}.
A larger block size will increase the signal-to-noise ratio per block, allowing the variability in $P_3$ to be better determined via the cross-correlation explained below. However, at the same time, a larger block size would cause smearing since the fixed period folding within the block might no longer be accurate enough.
The variations in $P_3$ were taken into account by cross-correlating the folded data for each block, resulting in phase offsets, which are applied to align the average drift bands obtained for the separate blocks. This is done in an iterative manner. In the first iteration, the first block of individual pulses is used as the template in the process of cross-correlation when aligning the other blocks. In the next iteration, the over the full data set averaged modulation cycle from the previous iteration is used as the template, thus giving a better alignment.
The number of iterations was chosen such that further iterations do not lead to  significant improvement in the final result.

The calculated offsets from the cross-correlation can be strongly influenced by the presence of randomly occurring bright pulses weakening the correlation of the  underlying drifting subpulses pattern. For example, \citet{kss11} showed that for PSR~B0031$-$07 the emission is a combination of drifting subpulses, nulls and sporadic bright pulses. In order to minimise the mis-alignment because of sporadic bright non-periodic emission, all samples above a certain selected threshold of intensity were clipped before folding. This means that their intensity was set to the threshold intensity when exceeding the threshold, thereby maximising the power in the drifting subpulses. This process is similar to that applied to Fig. \ref{fig:stack} to highlight the drifting subpulses better.
The offsets determined from the clipped data were used to fold the original not clipped pulse-stack. The data corresponding to the four Stokes parameters were folded identically using the offsets determined from the Stokes $I$ pulse-stack, in which the drifting subpulses are clearest. After folding, parameters such as $L$, PA and $\chi$ can be determined as usual. 

The $P_3$-folds in Stokes $I$, $L/I$, PA, $V/I$ and $\chi$ for both drift modes are shown in Fig.~\ref{fig:p3fold_modeA}. The average drift bands are shown twice, above each other, in the individual panels for the purpose of continuity. Looking at the Stokes $I$ $P_3$-folds, shown in the top row of $P_3$-folds in Fig.~\ref{fig:p3fold_modeA}, clear diagonal drift bands can be seen, once more indicating that the mode separation was done correctly.
The mode A drift bands (left) appear to be mostly straight diagonal bands with an intensity which is more concentrated to the central region of the profile compared to the mode B drift bands (right).

In the PA $P_3$-folds (fourth row of panels from the top in Fig.~\ref{fig:p3fold_modeA}) OPM$_\mathrm{A}$ (red in the online version) and OPM$_\mathrm{B}$ (blue in the online version) appear at different phases of the modulation cycle. 
This implies that throughout the modulation cycle, for both drift modes, the dominance of the two OPMs switches periodically.
For drift mode A (left) it can be seen that between pulse longitudes $170\degr$ and $182\degr$, OPM$_\mathrm{A}$ roughly coincides with the location of the total intensity drift band (represented by the contours), and OPM$_\mathrm{B}$ occurs mostly in between the total intensity drift bands. After pulse longitude $182\degr$, an inversion occurs in the polarization such that the total intensity drift band is dominated by OPM$_\mathrm{B}$, while OPM$_\mathrm{A}$ occurs in between the total intensity drift bands. 
At all pulse longitudes where in the PA distribution (Fig.~\ref{fig:PA_all}) the dominance of OPMs alternates,
the transitions are part of the modulation cycle. 
Also the drift mode B data (right hand side of Fig.~\ref{fig:p3fold_modeA}) show periodic switches between the two OPMs after pulse longitude $177\degr$, although here the total intensity drift bands are dominated by OPM$_\mathrm{A}$ rather than OPM$_\mathrm{B}$.
At earlier longitudes, OPM$_\mathrm{B}$ dominates at all times (see also the PA distribution in Fig.~\ref{fig:PA_all}), hence no vertical structure is seen in the $P_3$-fold.

In the $P_3$-fold of $L/I$ for drift mode A, shown below the Stokes $I$ $P_3$-fold in Fig.~\ref{fig:p3fold_modeA}, it can be seen that the fractional linear polarization drops  where the PA switching occurs (compare with the second but last left panel). Depolarization is expected to occur where OPMs mix.
A similar behaviour can be seen in the $L/I$ $P_3$-fold of drift mode B for that part of the profile where  OPM switching is observed.

In the $P_3$-folds in $V/I$, displayed in the second to last row of panels in Fig.~\ref{fig:p3fold_modeA}, it can be seen that the circular polarization is weakly modulated by the periodicity of drifting subpulses. Given that OPM$_\mathrm{A}$ is elliptical, we expect Stokes $V$ to be modulated as well. Indeed we see that for mode A (left hand side), between pulse longitudes $170\degr$ and $182\degr$, periodic switches occur between a significant circular polarization component (coinciding with OPM$_\mathrm{A}$) and almost no circular polarization (coinciding with OPM$_\mathrm{B}$). A similar trend can be seen for mode B after pulse longitude $170\degr$.

In the bottom left panel, where the $P_3$-fold of $\chi$ of drift mode A is shown, one can see that OPM$_\mathrm{A}$ has significant ellipticity, while OPM$_\mathrm{B}$ is almost not elliptical. 
Particularly interesting is an asymmetry in time (pulse number) in the ellipticity for especially drift mode A. 
For that mode, between pulse longitude $170\degr$ and $180\degr$, the drift band corresponding to OPM$_\mathrm{A}$ starts with a lower ellipticity and it becomes highly elliptical when the OPM flips and the fractional linear polarization drops, as indicated by the (in the online version) red edge corresponding to the upper side of OPM$_\mathrm{A}$ in the bottom-left panel of Fig.~\ref{fig:p3fold_modeA}.
For the same drift mode, between pulse longitudes $184\degr$ and $194\degr$,  OPM$_\mathrm{A}$ starts highly elliptical and becomes more linear during the modulation cycle.
An asymmetry is also present for drift mode B (bottom right panel), however the effect is less pronounced as for mode A.
Up to pulse longitude $185\degr$, the ellipticity of the OPM$_\mathrm{A}$ drift band is symmetrical in pulse number, and is maximum when the OPMs flip (bright red edge in the online version), however after this longitude only the start of the OPM$_\mathrm{A}$ band has a hint of ellipticity. 
For both drift modes, asymmetries in pulse number are only apparent in ellipticity.

\section{Discussion}
\label{sect:discussion3}

We have discovered that for PSR~B0031$-$07 the switches between the two OPMs are modulated synchronously with the drifting subpulses for both the A and B drift modes. This phenomenon has only been seen for four other pulsars so far, and none of those pulsars have multiple stable drift modes.

\citet{rr03} suggested that this behaviour can be accommodated in the carousel model. In this model, drifting subpulses reflect a pattern of sub-beams circulating around the magnetic pole and the origin of this pattern is ``sparking'' (pair production) near the surface of the star. To explain the observed behaviour two patterns, or ``images'' are required, each corresponding to one of the OPMs. These two images are thought to naturally arise because of birefringence in the pulsar magnetosphere. 
Although at the emission height, the height where the two OPMs are produced, the ray paths corresponding to the two OPMs are identical, birefringence will cause the rays to escape the magnetosphere in different directions. As a consequence, the observed beam patterns corresponding to the OPMs differ, and provide separate distorted views (``images'') of the same underlying pattern of sparks.
To explain their observations, \citet{rr03} found that these images should be shifted both radially and azimuthally with respect to each other. Thus for PSR~B0031$-07$, this ``double imaging'' should happen for (at least) both the A and B modes of drifting, although we will point out that this picture must be more complicated for this pulsar.

\citet{smk05,sms+07} constructed a geometrical model in order to explain the drifting subpulses (as observed in total intensity) of PSR~B0031$-07$. The authors suggested that the emission of this pulsar comes from a carousel configuration which expands or contract depending on the drift mode. The widest carousel configuration would correspond to drift mode A. 
This model can explain the fact that the intensity and occurrence of mode B drift bands decreases with increasing observing frequency, while the occurrence of mode A increases with observing frequency.  This model relies on radius-to-frequency mapping (RFM; \citealt{cor78}), which is the effect where higher frequency emission originates from closer to the stellar surface, from magnetic field lines which are further away from the magnetic axis.

The \citet{smk05,sms+07} model was not designed to explain the polarization properties of this pulsar. 
In the context of a symmetrical circular carousel, such as was applicable to PSR~B0809+74 at 328~MHz \citep{rrv+06}, one would expect to observe a symmetry of the drift bands with respect to the centre of the profile. We do not see such symmetry in the  $P_3$-folds of PSR~B0031$-$07, as an inversion occurs in the PA of drift mode A (see Sect.~\ref{sect:p3fold})  such that the drift band as seen in total intensity is dominated by OPM$_\mathrm{A}$ in one half of the profile, and OPM$_\mathrm{B}$ in the other half. For mode B, in the earlier part of the pulse there is only OPM$_\mathrm{B}$, while in the trailing half of the profile we see switches of the OPM synchronised with the drifting subpulses. 
This asymmetry prevents the methodology of \cite{rr03} to be applied to construct a polarized beam map for PSR~B0031$-$07. However, it is clear that two offset images of the same carousel are not enough to explain our observations, and additional effects must play a role.
These could potentially be asymmetric propagation effects in the pulsar magnetosphere.
The same conclusion would apply for PSR~B0818$-$13, which has complicated, and asymmetric, structures in its polarization, which are synchronously modulated with the drifting subpulses. 
Although \cite{edw04} does not discuss the asymmetry of the drift bands specifically, it was concluded that for PSR B0818$-$13, which has two non-orthogonal highly elliptical OPMs, a model with two offset OPM images of the same carousel is not enough to explain the observed polarization behaviour.

The fact that the $P_3$-folds significantly differ for the two drift modes as observed in polarization suggests that more is happening as a simple expansion or contracting of a single carousel. Again, this could point to propagation effects to play a role, since for a changing carousel configuration the emission would propagate through different parts of the magnetosphere giving rise to different propagation effects for the different drift modes.

In the case of PSRs B0320+39 and B0818$-$13 \citep{edw04}, and also PSR~B0809+74 at both 328~MHz and 1380~MHz \citep{edw04,rrv+06}, the total intensity drift bands are dominated by one OPM, while the other OPM dominates in between drift bands. 
PSR~B0031$-$07 is different, since the total intensity drift band in drift mode A is dominated by different OPMs in the leading and trailing half of the profile.
Looking at the $P_3$-fold of drift mode A in PA (fourth row of panels from the top in Fig.~\ref{fig:p3fold_modeA}), the pattern traced by a given OPM appears to be split in the middle of the profile with a vertical offset appearing. This at least visually resembles a subpulse phase jump as seen, for example, in the total intensity drift bands of PSR~B0320+39 \citep{esv03}. For PSR~B0031$-$07 the total intensity $P_3$ folds are more continuous, although coincident with the discontinuity as seen in the PA $P_3$-fold of drift mode A, a small change in the slope of the intensity drift band can be seen, such that the drift band becomes shallower  after the discontinuity. \citet{esv03} suggested that a subpulse phase jump is another type of phenomenon which might require superposed images of the same carousel system. Here each image should have drifting subpulses which are out of phase because of an azimuthal offset. 
To explain the subpulse phase jump, each image should dominate in different halves of the profile. At the longitude where the phase modulation jump occurs, in the middle of the profile, the two images destructively interfere. \citet{edw04} pointed out that if the two superposed out of phase images  correspond to the two OPMs, both the subpulse phase step as seen in total intensity and the synchronous switching of the dominant OPM can be explained. 
\citet{es03} observed the intensity phase modulation of PSR~B0320+39 at 328 and 1380~MHz and noted that the change in the phase jump with frequency could be explained if the centre of the carousel systems are offset from each other. So there would be a precedent to explore if a model which is more complicated than just a shrinking or expanding carousel would apply to PSR B0031--07 as well.

Other authors (e.g. \citealt{cr04,cr08,rd11}) suggested that non-radial oscillations of the surface of the neutron star could explain drifting subpulses as well as the subpulse phase steps observed in total intensity drift bands.
However, it is unclear in this model how the polarization is expected to be affected, and why for PSR~B0031$-$07 no subpulse phase step would be observed in total intensity, while there is a discontinuity in the PA of the drift bands in the middle of the profile.

It is intriguing that the ellipticity evolves asymmetrically throughout the modulation cycle (in pulse number), such that in drift mode A in the leading half of the profile, the emission becomes elliptical just before the emission flips from OPM$_\mathrm{A}$ (which as found in Sect.~\ref{sect:polarization} is significantly more elliptical on average) to OPM$_\mathrm{B}$, but not when OPM$_\mathrm{B}$ flips to OPM$_\mathrm{A}$.
Hence, the ellipticity of the OPMs are not fixed during the modulation cycle.
In the carousel model, the pattern of circulating sub-beams is in general taken to be symmetric with respect to the sense of circulation, in the sense that the structure of the sub-beam is similar in its leading and trailing half. Here, some form of symmetry breaking is required by physics which is affected by the sense of the circulation. 
Similar symmetry breaking has not been seen for the other four pulsars which show  OPM switches synchronous with drifting subpulses.
The origin of this asymmetry is unclear, but coherent interference by the two OPMs could play a role.

From theory (e.g. \citealt{ab86}) one could expect the OPMs to be purely linearly polarized. However, this is not the case for PSR~B0031$-$07. 
Also PSR~B0329+54 has two OPMs with different degrees of ellipticity \citep{es04}. The authors explained this in term of Generalised Faraday Rotation (GFR, \citealt{km98}) and coherent addition of the radiation of the two OPMs, which allows partial conversion of linear into circular polarization. GFR would contribute to frequency dependend polarization effects as well (see \cite{ijw19} and references therein).
\citet{mmk+06} offered an alternative explanation for the observed variable polarization properties of individual pulses of PSR~B0329+54, by assuming incoherent addition of the radiation of the two OPMs.
In general it is hard to distinguish whether the observed polarized radiation is the incoherent or coherent sum of OPMs.
For PSR~B0031$-$07, when the rate of occurrence of both OPMs is comparable in the PA distribution of the not drift-mode separated dataset, the linear polarization is generally low, while when the rate of occurrence of one OPM is higher, as seen for the drift mode A separated dataset, $L$ is also higher. This is compatible with a scenario where the observed radiation is an incoherent sum of OPMs. On the other hand, as pointed out for example by \citet{dyk17}, a good indicator for the natural polarization modes being coherently combined would be the presence of a peak of circular polarization near an OPM transition, while $L/I$ decreases and the total polarization is not affected.
Fig.~\ref{fig:PA_all} shows that the ellipticity angle peaks where the OPM transitions occur for drift more A (longitudes $171\degr$ and $178\degr$), however the average total polarization does significantly drop. So it seems likely there is a combination of coherent and incoherent OPM summation (partial coherent addition) occurring in PSR~B0031--07.

\section{Summary and Conclusions}
\label{sect:conclusions3}

We established, using the $P_3$-folding technique, that for PSR~B0031$-$07 the orthogonal polarization modes switch synchronously with the drifting subpulses seen in total intensity. This is observed for both drift modes (out of a total of three) which are present at our observing frequency \citep{smk05,sms+07}. 
A similar connection between the periodicity seen in polarization and total intensity has only been reported for four other pulsars \citep{rrs+02,rr03,edw04}. PSR~B0031$-$07 is the only pulsar in this group of pulsars which shows drift mode changes, making it a unique source to test emission models.

PSR~B0031$-$07 is also different from the other four pulsars, as it was found  that the ellipticity evolves asymmetrically with time during the modulation cycle, such that 
the ellipticity is high just before the emission flips from one OPM to the other, but not when it flips back. The asymmetry is different for the two drift modes.
In the carousel model, some form of symmetry breaking would be required by physics which is affected by the sense of the circulation, which is likely related to coherent mode addition effects.

For all other four pulsars which show modulated OPM switches synchronous with drifting subpulses, the drift bands observed in total intensity are generally dominated by one OPM, while the other one dominates in between the drift bands. 
However, during drift mode A of PSR~B0031$-$07, the drift bands as observed in total intensity are dominated by different OPMs in different halves of the profile.
This is accompanied by a slight change of the slope of the intensity drift band in the middle of the profile. 
This could be related to what is known as subpulse phase steps \citep{esv03}
and could hint towards two superposed out of phase images of the carousel.
The production of two images is also a requirement to explain the observed periodic changes in polarization.

It will be challenging to construct a geometric model for the polarized properties of the drifting subpulses of PSR~B0031$-$07. 
A symmetrical carousel which contracts or expands when switching between drift modes, as was suggested by \citet{smk05,sms+07} for PSR~B0031$-$07, can only explain the behaviour as seen in total intensity. Additional effects would be required in order to explain the here presented polarization observations which show large asymmetries between the leading and trailing halves of the profile. 
This is likely to be because of propagation effects in the pulsar magnetosphere, including birefringence which can cause the emission produced by the carousel to be spatially split in two distinct images corresponding to two OPMs. In addition, it is suggested that a combination of coherent and incoherent mode addition plays a role in forming the observed polarization state of the radiation. To explain the large asymmetries as seen in the leading and trailing half of the profile, especially the $P_3$-fold as observed in PA, we speculate that more complex propagation effects play a role, such as the coupling between the two OPMs during the propagation through the magnetosphere. If this coupling is asymmetric as function of pulse longitude, one could expect asymmetries to be observed, while maintaining the picture of a symmetric underlying carousel configuration.

\section*{Acknowledgements}

The authors would like to thank Ben Shaw for providing the ephemeris used in the folding process. In addition we would like to thank Crispin Agar and Geoff Wright for the many useful discussions related to the interpretation of the results in this manuscript. The Parkes radio telescope is part of the Australia Telescope which is funded by the Commonwealth of Australia for operation as a National Facility managed by CSIRO. Pulsar research at Jodrell Bank Centre for Astrophysics and Jodrell Bank Observatory is supported by a consolidated grant from the UK Science and Technology Facilities Council (STFC). 




\bibliographystyle{mnras}
\bibliography{ms.bib} 








\bsp	
\label{lastpage}
\end{document}